\documentclass{article} 

\usepackage{graphicx}
\graphicspath{{converted_graphics/}}
\begin{document}

\begin{center}
{\LARGE Crystal Growth in the Presence of Surface Melting and Impurities: An Explanation of Snow Crystal Growth Morphologies}%
\vskip16pt

{\Large Kenneth G. Libbrecht}\footnote{%
e-mail address: kgl@caltech.edu\newline
For the latest version of this paper, go to
http://www.its.caltech.edu/\symbol{126}atomic/publist/kglpub.htm}{\Large \ }%
\vskip4pt

{\large Department of Physics, California Institute of Technology}\vskip-1pt

{\large Pasadena, California 91125}\vskip-1pt

\vskip18pt

\hrule \vskip1pt \hrule
\vskip 14pt
\end{center}

\noindent \textbf{Abstract. We examine the molecular dynamics of crystal
growth in the presence of surface melting and surface impurities, and from
this propose a detailed microscopic model for the growth of ice from the
vapor phase. Our model naturally accounts for many aspects of the
experimental data that are otherwise difficult to explain, and it suggests a
variety of measurements that can provide further confirmation. Although
additional work is needed to refine these ideas, we believe that the
combined influences of surface melting and impurities provide a viable
solution to the 60-year-old mystery of why snow crystal morphologies vary so
dramatically with temperature. }

\section{\noindent Introduction}

The formation of complex structures during solidification often results from
a subtle interplay of nonequilibrium, nonlinear processes, for which
seemingly small changes in molecular dynamics at the nanoscale can produce
large morphological changes at all scales. One popular example of this
phenomenon is the formation of snow crystals, which are ice crystals that
grow from water vapor in an inert background gas. Although this is a
relatively simple physical system, snow crystals display a remarkable
variety of columnar and plate-like forms, and much of the phenomenology of
their growth remains poorly understood \cite{libbrechtreview}.

\begin{figure}[t] 
  \centering
  \includegraphics[width=5.5in,keepaspectratio]{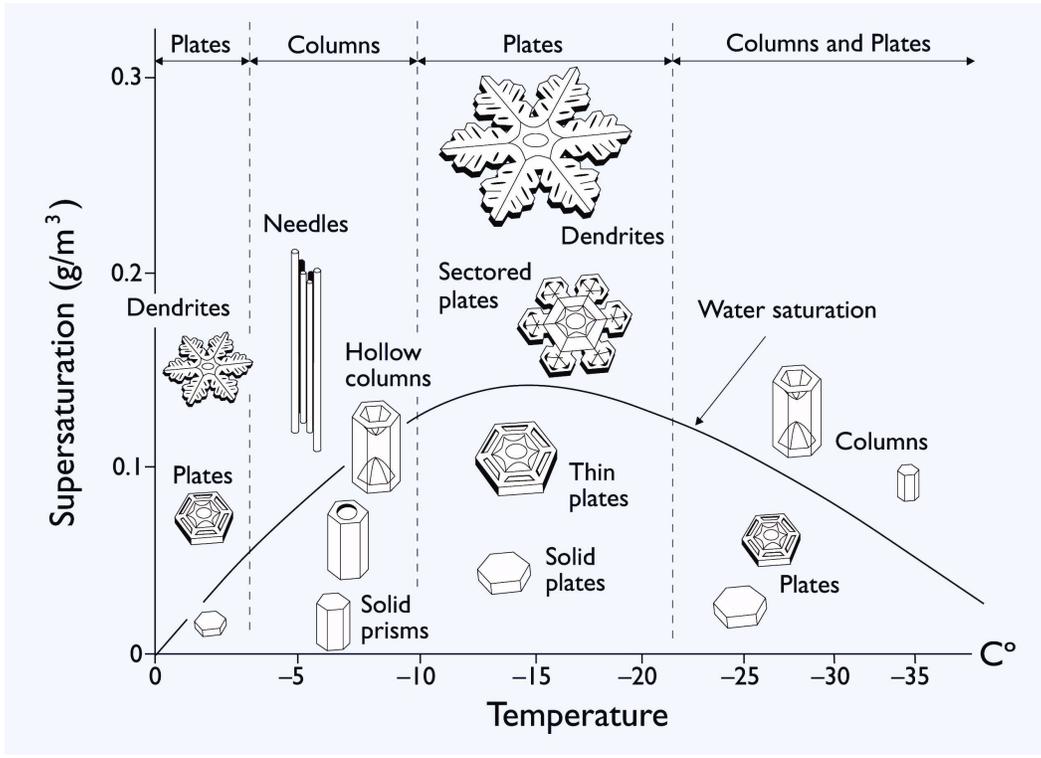}
  \caption{The \textit{snow crystal morphology diagram}, showing how the morphology of
ice crystals growing from water vapor in air at a pressure of one bar
changes with temperature and supersaturation. The solid line indicates the
supersaturation of supercooled water relative to ice. At high
supersaturations, the overall structure alternates between plate-like and
columnar forms as the temperature is reduced.}
  \label{morph}
\end{figure}

One of the most enduring puzzles surrounding snow crystal formation has to
do with changes in the growth morphology with temperature. As shown in
Figure \ref{morph}, the overall structure of snow crystals grown in air
alternates twice between plate-like and columnar forms as a function of
temperature. This behavior was first observed in the early 1940s \cite%
{nakaya}, and for over 60 years scientists have been unable to provide even
a qualitative explanation for why snow crystal morphologies exhibit this
temperature dependence \cite{libbrechtreview}. Mason \cite{mason, mason2}
suggested that the basic habit is determined by variations in surface
diffusion rates on the basal and prism facets, but did not provide an
explanation of why the rates alternated with temperature in the proposed
manner. Nelson and Knight \cite{nelson1, nelson2} suggested that
morphologies are determined by variations in layer nucleation rates on the
basal and prism facets, but again did not explain why the rates alternated
with temperature. The Lacmann-Stranski-Kuroda (LSK) model \cite{lacmann1972,
kuroda1982, kobayashi1987} was perhaps the most ambitious attempt to create
a microscopic model of ice crystal growth. In this model, the authors
proposed that variations in growth rates on the two facet surfaces are
brought about by temperature-dependent changes in the structure of the ice
surface related to surface melting. As we will see below, however, the LSK
model does not agree with recent ice growth data.

\begin{figure}[t] 
  \centering
  \includegraphics[width=4.5in,height=3.12in,keepaspectratio]{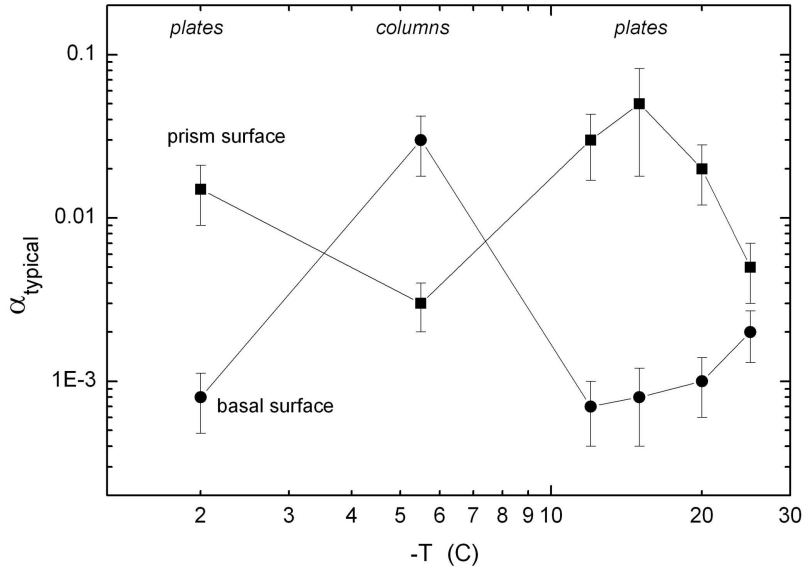}
  \caption{Ice
growth measurements in ordinary laboratory air at one bar, for which
diffusion modeling has produced $\protect\alpha $ for the basal and prism
facets \protect\cite{growth2008}. Here we show data for supersaturations
typical in our experiments, ranging from $\protect\sigma _{\infty }\approx
1.5$ percent at $T=-2$ C to $\protect\sigma _{\infty }\approx 3$ percent at $%
T=-20$ C. For the small crystals observed, $\protect\sigma _{surf}$ was
typically no less than $\protect\sigma _{\infty }/2$ over the entire crystal
surface. Note that these data show a dependence with temperature that is
expected from the morphology diagram.}
  \label{alphacomb}
\end{figure}

In this paper, we examine how surface melting and surface impurities can
affect crystal growth, and from this propose a new microscopic model for ice
crystal growth as a function of temperature. We also propose a new
impurity-mediated growth instability that promotes the formation of thin
plate-like crystals as well as slender columns or sheath-like structures.
Our model suggests that impurities play an essential role in determining
snow crystal growth morphologies under normal atmospheric conditions, and
that growth in perfectly clean air would be markedly different from that
shown in Figure \ref{morph}. While the effects of surface impurities have
been largely ignored in previous work, our model suggests that impurities
have profoundly affected essentially all experimental observations to date.
Finally, we comment on how snow crystal growth morphology should vary with
impurity level, and we propose new experiments that can provide further
insights into this phenomenon.

\section{Developing a Picture of Ice Crystal Growth}

\subsection{Basic Considerations}

For an ice crystal growing from water vapor, we can write the growth
velocity normal to the surface in terms of the Hertz-Knudsen formula \cite%
{libbrechtreview, saito} 
\begin{equation}
v_{n}=\alpha v_{kin}\sigma _{surf}  \nonumber
\end{equation}%
where $v_{kin}$ is a temperature-dependent kinetic velocity derived from
statistical mechanics, $\sigma _{surf}\ $is the water vapor supersaturation
just above the growing surface, and $\alpha $ is the condensation
coefficient, which contains the surface physics that governs how water
molecules are incorporated into the ice lattice, collectively known as the
attachment kinetics \cite{libbrechtreview}. The attachment kinetics
describing ice crystal growth is quite complex, so in general $\alpha $ will
depend on $T,$ $\sigma _{surf},$ the surface structure and geometry, surface
chemistry, etc. If molecules striking the surface are immediately
incorporated into it, then $\alpha =1;$ otherwise $\alpha \leq 1$.

If an ice crystal is growing in an inert background gas, then diffusion of
water molecules to the ice surface often significantly impedes the growth 
\cite{libbrechtreview}, so that $\sigma _{surf}<\sigma _{\infty }.$ The
physics of the diffusion process is well known, and computer models can be
used to calculate growth rates and morphologies from known attachment
kinetics, or to extract $\alpha $ values from growth measurements \cite{gg,
libbrechtmodel}. Laboratory measurements of the growth of small ice crystals
in air at one bar, in conditions similar to those found in the atmosphere,
allowed us to extract the typical values of $\alpha $ shown in Figure \ref%
{alphacomb} \cite{growth2008}. Note that the changes in $\alpha _{basal}$
and $\alpha _{prism}$ with temperature in this figure agree with the
temperature dependence seen in the morphology diagram, as one would expect.

In what follows we develop a microscope model for $\alpha $ under conditions
that are typical for atmospheric ice growth and other conditions relevant
for laboratory experiments.

\subsection{Surface Melting}

Surface melting refers to an equilibrium phenomenon in which a quasi-liquid
layer (QLL) exists at the surface of a crystalline solid near the melting
point. The thickness of the QLL is generally a strong function of
temperature, diverging as the bulk melting point is approached. As the name
suggests, both the structure and molecular dynamics of the QLL resemble that
of a fully melted liquid, at least when the layer is sufficiently thick.
Surface melting is present in most crystalline solids \cite{dash}, and it
has been especially well studied for ice \cite{petrenko, wet1}. Because
surface melting is central to our ice growth model, we summarize the basic
concepts here.

One useful way to think about surface melting is from the perspective of the
Lindemann criterion \cite{linde, libbrechtreview}. This century-old idea
states simply that a crystalline solid will melt when thermal motion of
molecules in the crystal lattice produce excursions from equilibrium that
are greater than approximately 10-15\% of the normal lattice spacing. The
Lindemann criterion works reasonably well for a wide range of bulk materials 
\cite{linde2}, and it has been directly verified both experimentally \cite%
{linde3} and in molecular dynamics simulations \cite{linde4}.

If we assume a simple model in which a molecule is held in a harmonic
potential, then thermal motion produces average oscillatory amplitudes of
approximately 
\[
x=\left( \frac{kT}{\kappa }\right) ^{1/2} 
\]%
where $\kappa $ is an effective spring constant. Near the solid surface the
molecular binding is weaker than in the bulk, so we write 
\[
\kappa (d)=\kappa _{0}-f(d)\Delta \kappa 
\]%
where $d$ is the distance from the surface, $0<\Delta \kappa <\kappa _{0}$
is a constant, and $f(d)$ is a function with $f(0)=1$ and $f(\infty )=0.$
For the simplest model in which half the restoring force is absent at the
surface, we might expect $\Delta \kappa \approx \kappa _{0}/2.$

The Lindemann criterion states that the bulk will melt when $x$ reaches some
value $x_{0}=(kT_{m}/\kappa _{0})^{1/2},$ where $T_{m}$ is the bulk melting
temperature. Extending this, we expect that surface melting will occur down
to a depth $d$ at which 
\[
x(d)=\left( \frac{kT}{\kappa _{0}-f(d)\Delta \kappa }\right) ^{1/2}=x_{0}.
\]%
Expanding for small $\Delta \kappa /\kappa _{0}$ and for small $%
t=(T_{m}-T)/T_{m}$, this reduces to simply 
\[
t=f(d)\Delta \kappa /\kappa _{0}.
\]%
For the case of ice we assume $f(d)=\exp (-d/\lambda ),$ where $\lambda $ is
a scaling length for the intermolecular interactions \cite{dash}, and with
this the thickness of the quasi-liquid layer becomes 
\[
d(T)\approx \lambda \log \left( \frac{\Delta \kappa }{t\kappa _{0}}\right) 
\]%
which is valid for small $t.$

There have been numerous measurements of surface melting in ice \cite%
{petrenko}, and there is general consensus that the QLL thickness $d(T)$ at
least roughly follows the logarithmic relation described above. There is,
however, great disparity between the different measurements, probably
because most experiments do not measure $d(T)$ directly, but rather measure
some quantity related to $d(T)$ in an uncertain way (see \cite{dosch1995,
wei2002} and references therein for a review of recent measurements).
Molecular dynamics simulations are making progress toward understanding the
surface structure of ice, including surface melting \cite{bolton2000,
furukawa1997, kroes1992}, although reliable quantitative results are
elusive. As a result of both experimental and theoretical uncertainties, we
do not yet fully understand the equilibrium structure of the ice surface in
detail, or how this structure varies with temperature. Further, we do not
know if surface melting is different on different crystal facets, or in
general how the QLL varies with the surface orientation relative to the
underlying ice lattice.

In spite of these problems, theory and experiment do give us a relatively
simple qualitative picture of the development of surface melting in ice as a
function of temperature. At very low temperatures, surface melting is
largely absent, so there is relatively little disorder or time-dependent
restructuring at the surface. As the temperature increases, surface melting
begins turning on gradually, causing some disordered motion in the top few
molecular layers, and this motion increases in extent with increasing
temperature. Eventually this disordered motion starts to look like a fully
developed QLL that sits atop the bulk crystal, and a somewhat distinct
boundary appears between the two. Very close to the melting point, the QLL
is thick enough that it becomes an essentially distinct layer with largely
liquid-like properties. Finally, the layer thickness diverges as the melting
temperature is approached.

As was pointed out in connection with the LSK model \cite{lacmann1972,
kuroda1982, kobayashi1987}, it appears likely that surface melting plays an
important role in ice crystal growth dynamics. The available data indicate
that surface melting shows considerable variation over the temperature range 
$-20$ C $<T<0$ C, coincident with the most pronounced variations in snow
crystal morphology, and it is plausible that surface melting is
substantially different on the two principal facets over this range.
Unfortunately, we have little theoretical or experimental guidance as to how
the structural changes associated with surface melting affect crystal growth
dynamics \cite{bienfait}.

\subsection{The Role of Surface Impurities}

It has long been known that high levels of gaseous impurities in air at one
bar can dramatically affect snow crystal growth and morphologies (\cite%
{libbrechtreview} and references therein). A wide range of chemical
impurities, such as vapors from many alcohols, acids, hydrocarbons, etc.,
affect both ice crystal morphologies and growth rates. However, observations
over many decades have shown that the morphology diagram is a robust feature
of snow crystal growth in air, provided the air is fairly clean. Our own
\textquotedblleft rule of thumb\textquotedblright\ in the lab has been that
growth morphologies are largely unaffected as long as one cannot smell
anything in the air. Since impurity effects seen at high concentrations are
not seen in cleaner air, researchers were led to believe that the modest
levels of residual impurities in ordinary laboratory air or in the
atmosphere were probably not playing a significant role in ice growth
dynamics.

These observations may be deceptive, however, because even a seemingly low
concentration of surface-active impurities may not be negligible. Consider
growth in air which contains a fraction $f$ of some impurity that
immediately attaches to the ice surface upon contact. Diffusion limits the
transport of impurity molecules to the surface, and the time needed to form
a single monolayer on a spherical crystal is%
\begin{equation}
\tau _{mono}\approx \frac{R}{Da^{2}fn_{air}}  \label{mono}
\end{equation}%
where $R$ is the crystal radius, $D$ is the diffusion constant $(D=2\times
10^{-5}$ m$^{2}/\sec $ in air at one bar), $n_{air}$ is the number density
of the air $(n_{air}=2.5\times 10^{25}$ /m$^{3}$ at one atmosphere), and $a$
is the effective size of an impurity molecule (which we assume to be $%
a\approx 1$ nm). Equation \ref{mono} holds for times longer than the
characteristic diffusion time $\tau _{diff}=R^{2}/D$. Note that $D\sim
P^{-1} $ and $n_{air}\sim P,$ where $P$ is the air pressure, so $\tau
_{mono} $ is independent of $P$ to lowest order. This expression becomes%
\begin{equation}
\tau _{mono}\approx 2\left( \frac{R}{1\ \mu m}\right) \left( 
\frac{1\ ppm}{f}\right) \ msec  \label{taumono}
\end{equation}%
which is much shorter than typical snow crystal growth times. The growth
velocities of micron-scale ice crystals in air are typically no higher than
a few microns per second, so impurities may be playing a role even for very
small ice crystals.

For times much longer than $\tau _{mono}$ we expect that the impurity
density on the ice surface will reach some equilibrium value, assuming that
thermal excitations are sufficient to remove adsorbed molecules from the
surface at some rate. Gaseous impurities striking the surface will increase
the surface density as $d\rho _{surf}/dt\sim fn_{air},$ while losses give $%
d\rho _{surf}/dt=\rho _{surf}/\tau _{loss},$ where $\tau _{loss}$ is the
characteristic time for impurities to leave the surface. These rates are
equal in equilibrium, giving $\rho _{surf}\sim fn_{air}\tau _{loss}.$ Thus
while $\tau _{mono}$ is independent of $P$, we see that reducing the
pressure will reduce the equilibrium impurity surface density as $\rho
_{surf}\sim P$ (assuming $\tau _{loss}$ is independent of $P).$

Experience with other condensed matter systems also suggests that impurity
effects may be important. Creating atomically clean semiconductor surfaces
often requires extensive baking, sputtering, or other types of surface
cleaning, all of which must be done under ultrahigh vacuum conditions. The
high vapor pressure of ice near the melting point precludes the use of
ultrahigh vacuum techniques, and nearly all experiments with ice have not
involved careful surface preparation to remove adsorbed impurities.

\begin{figure}[t] 
  \centering
  \includegraphics[width=2.5in,keepaspectratio]{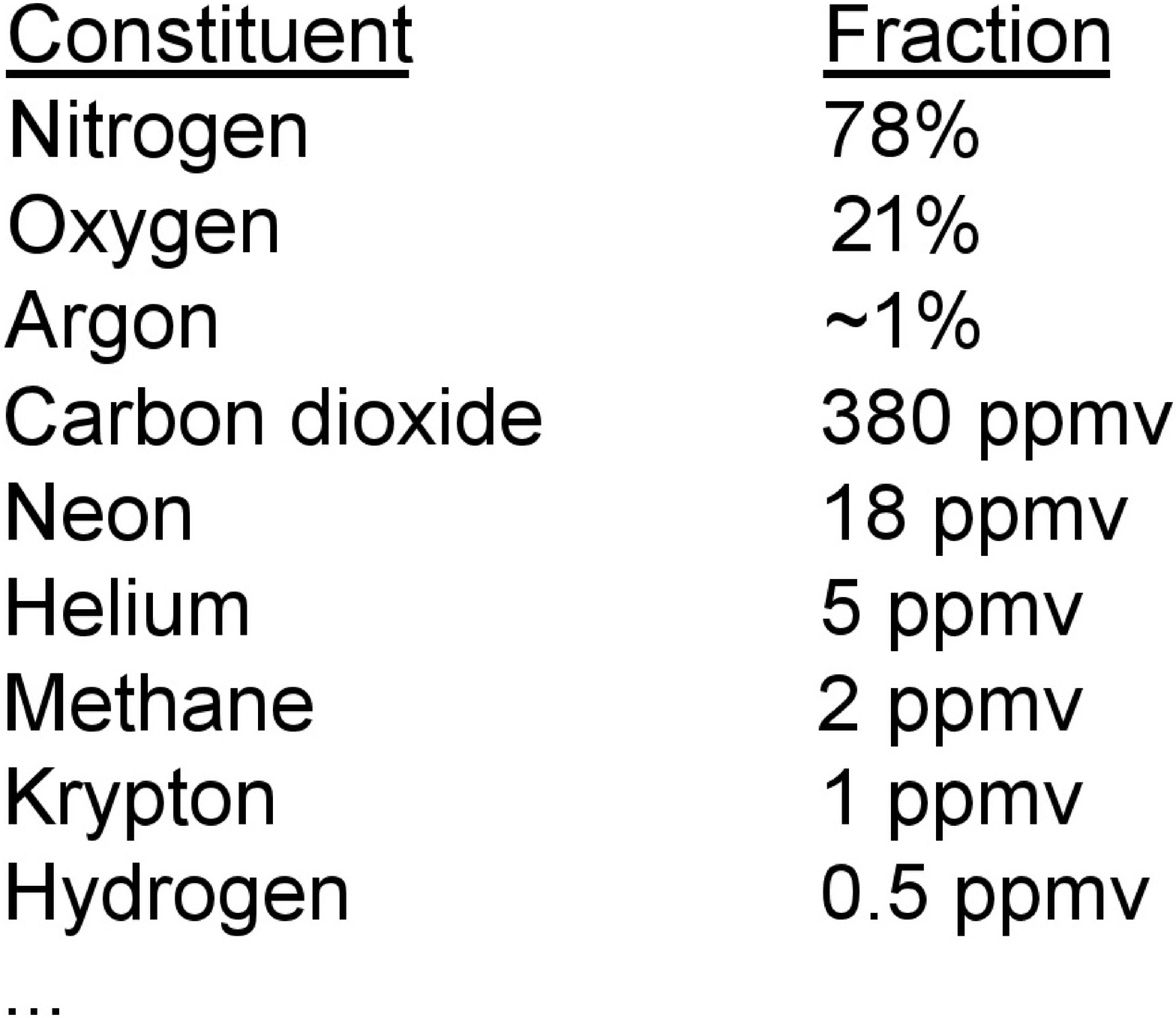}
  \caption{Constituents of clean dry air
under typical atmospheric conditions. Laboratory air, particularly in
chambers used for ice crystal growth experiments, is likely to contain
substantial concentrations of additional chemical impurities from a variety
of sources.}
  \label{impurities}
\end{figure}

Figure \ref{impurities} shows the typical composition of clean dry air. We
expect that laboratory air, particularly inside closed chambers used for ice
crystal growth experiments, would contain a long list of solvent vapors,
volatile organic compounds and other chemical impurities near or above the
ppm level.

With these simple considerations, we are led to suspect that even low levels
of impurities in relatively clean air could strongly affect ice crystal
growth. The remainder of this paper examines issues of impurity residence
time, mobility, how these change in the presence of surface melting, and
generally how impurities affect ice crystal growth rates and morphologies.

\subsection{Growth of a Clean Ice Surface}

Crystal growth involves complex dynamical processes at the molecular level,
and the task of modeling this phenomenon is certainly made more difficult
with the addition of surface melting and surface impurities. Our relatively
poor theoretical understanding of crystal growth in these circumstances
makes it difficult to develop an accurate \textit{a priori} model, so we use
experimental observations as much as possible to guide our thinking. For our
baseline model of the growth of a clean ice surface, we rely mainly on
growth measurements taken at low background pressure \cite{principal,
precision}. Not only is $\rho _{surf}$ expected to be lower for low-pressure
measurements, as described above, but we also believe gettering from the ice
reservoir in our growth chamber reduces the impurity fraction $f$ as well,
which in turn reduces $\rho _{surf}$. As we discuss further in the next
section, we believe these two factors greatly decrease the impurity effects
on our growth measurements. Based mainly on these low-pressure growth
measurements, we propose a model for the growth of an impurity-free ice
surface that includes the following properties:

\textbf{The surface attachment kinetics are described by 2D
nucleation-limited growth.} Crystal growth via 2D nucleation has been well
studied, and for such growth we can write $\alpha \approx A\exp (-\sigma
_{0}/\sigma _{surf})$ \cite{saito, libbrechtreview}, where the surface
diffusion parameter $A$ and the critical supersaturation $\sigma _{0}$ are
both functions of $T$, surface orientation, and potentially other
parameters. Our low-pressure measurements generally fit this functional form
well over the full range of our measurements \cite{principal, precision}. In
some cases we observed faster-growing crystals with $\alpha \sim \sigma
_{surf},$ indicating growth in the presence of spiral dislocations \cite%
{saito}, but these cases were relatively rare \cite{precision}.

\textbf{The critical supersaturation }$\mathbf{\sigma }_{\mathbf{0}}$\textbf{%
\ decreases smoothly and monotonically with increasing temperature, and is
essentially the same for both the prism and basal facets.} The justification
for this statement comes mainly from \cite{principal}, in which this
behavior was observed for $-40$ C $<T<-10$ C. From the theoretical side we
expect that surface melting will decrease the edge free energy $\beta $ of
an atomic step on the vapor/solid interface as the QLL thickens, until it
eventually becomes equal to $\beta $ for a step at the liquid/solid
interface. Given the gradual onset of surface melting described above, we
expect that $\beta $ (and thus $\sigma _{0}$) would decrease smoothly and
monotonically with increasing temperature, as we observed \cite{principal}.
This behavior is in contrast to the LSK model, which predicts large,
nonmonotonic changes in $\sigma _{0}$ with temperature.

\textbf{The }$\mathbf{A}$\textbf{\ factor is large for clean surfaces, with
little temperature dependence.} Recent observations \cite{hysteresis}
suggest that the measurements of $A$ in \cite{principal} were contaminated
to some degree by surface impurities, which reduced the measured growth even
at low pressure. We now believe that $A$ is generally quite large for
impurity-free surfaces \cite{precision, hysteresis}, with only a weak,
monotonic dependence on temperature, but this hypothesis is not well tested.

It appears that essentially all ice growth experiments to date have been
contaminated by surface impurities to some degree, so our clean-ice model is
quite uncertain. For considering the morphology diagram, however, the exact
growth behavior of a clean surface is less important than the fact that the
growth is fast. Our model suggests that ice growth in air under normal
atmospheric conditions is predominantly limited by surface impurity effects,
as we describe next.

\begin{figure}[t] 
  \centering
  \includegraphics[width=5in,keepaspectratio]{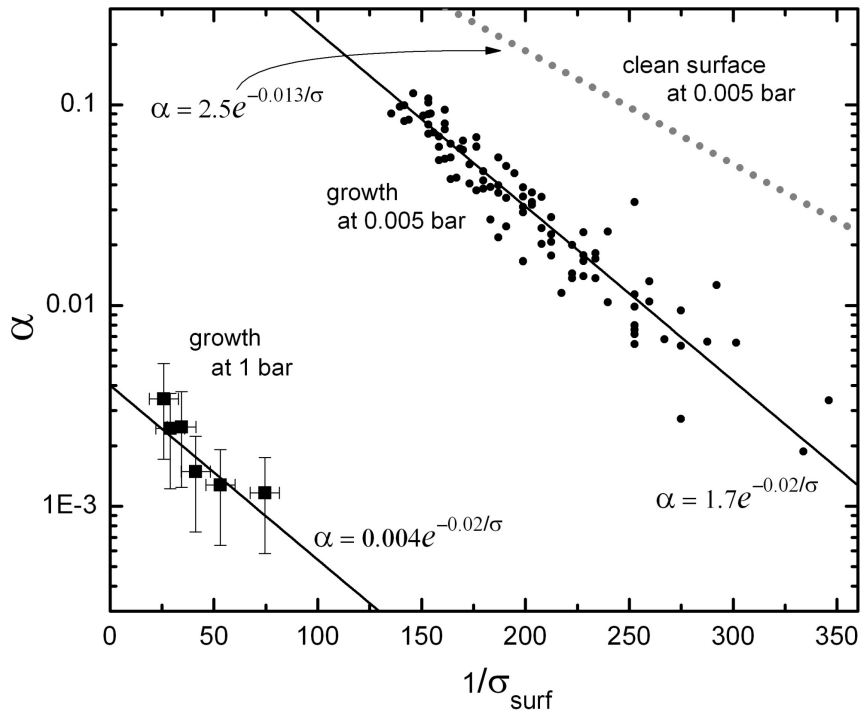}
  \caption{Measurements of the attachment
coefficient $\protect\alpha $ as a function of inverse surface
supersaturation $\protect\sigma _{surf}^{-1}$ for basal surfaces at -15 C,
comparing data taken at different background air pressures \protect\cite%
{precision, hysteresis, growth2008}. The top dotted line shows the rough
locus of our highest measured growth rates for ice surfaces that have been
cleaned as described in the text \protect\cite{hysteresis}, which we
interpret as growth of an impurity-free surface.}
  \label{lowp}
\end{figure}

\begin{figure}[ht] 
  \centering
  \includegraphics[width=3.5in,keepaspectratio]{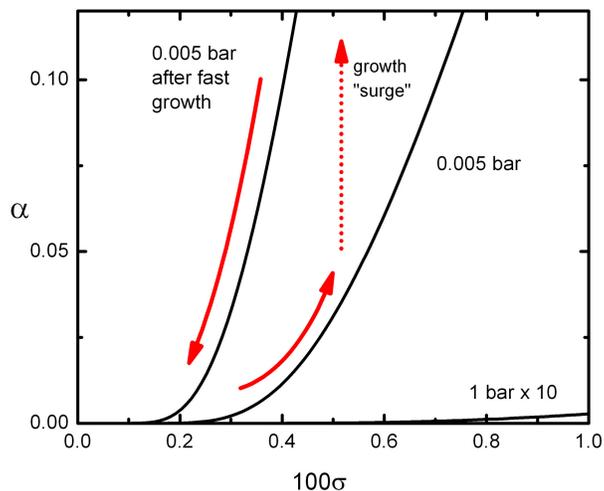}
  \caption{Schematic view of an unusual
hysteresis behavior seen in ice growth \protect\cite{hysteresis}, shown here
for the example of the basal surface at -15 C. Initial growth at a
background pressure of 0.005 bar is relatively slow (curve labeled 0.005
bar). As the supersaturation is slowly increased (lower arrow), the growth
suddenly exhibits a \textquotedblleft surge\textquotedblright , in which $%
\protect\alpha $ can suddenly increase by a factor of four or more. We
believe this event occurs because growth drives off or buries impurities on
the ice surface. Once the impurities have been removed, they do not reappear
quickly at low pressure, so $\protect\alpha (\protect\sigma )$ remains high
(upper curve) as $\protect\sigma $ is varied. If the surface impurity
density is allowed to build up again (by letting the crystal rest with $%
\protect\sigma =0$ for a few minutes), then the growth reverts to a lower $%
\protect\alpha (\protect\sigma ).$ At 1 bar, the surface impurity density is
always high, so $\protect\alpha $ is always very low (the lowest curve shows 
$10\protect\alpha $ for typical growth at 1 bar).}
  \label{theory}
\end{figure}

\subsection{Growth in the Presence of Surface Impurities}

The most basic and important feature of our new ice growth model is the role
of surface impurities. Whereas this factor has been largely ignored in
earlier models, we believe surface impurities can reduce growth rates by
orders of magnitude compared to growth of a clean surface, even in normal,
reasonably clean atmospheric conditions. Furthermore, we believe that the
temperature-dependent interplay of surface impurities with surface melting
is largely responsible for the most enigmatic features seen in the snow
crystal morphology diagram.

\subsubsection{Low Pressure versus High Pressure}

A major piece of evidence pointing to the importance of surface impurities
comes from a comparison of ice crystal growth rates in air at high and low
pressure, for which we show an example in Figure \ref{lowp}. These data show
the condensation coefficient $\alpha $ for the basal surface at -15 C, but
data at other temperatures and for the prism facet show similar trends. For
growth at a pressure of one bar, the effects of diffusion have been
subtracted by modeling \cite{libbrechtmodel}, giving $\alpha $ as a function
of $\sigma _{surf}$, the supersaturation at the growing surface. (For the
small crystals measured, $\sigma _{surf}$ and $\sigma _{\infty }$ typically
differed by no more than a factor of two, so modeling uncertainties do not
significantly affect the results shown.) Effects of diffusion are negligible
for growth at 0.005 bar.

Looking first at the two data sets in Figure \ref{lowp}, we see that the
growth rate (as parameterized by $\alpha $) at one bar is 2-3 orders of
magnitude smaller than at 0.005 bar. This behavior is inconsistent with all
the previous ice growth models described above, which predict no substantial
changes in $\alpha $ with pressure. Since diffusion has already been
accounted for in Figure \ref{lowp}, we must ask what other physical effects
could cause such a large difference in growth with pressure.

We feel it is unlikely that simple pressure forces can explain these data,
since the air density is low enough that pressure effects should add a
negligible perturbation to the surface molecular forces that determine the
attachment kinetics. We examined whether the major chemical constituents in
air could be responsible by measuring crystal growth rates in air, nitrogen,
and argon at one bar \cite{growth2008}. To the accuracy of our data (better
than a factor of two), all three gases yielded growth rates identical to
those shown in Figure \ref{lowp}. These data suggest that the major
atmospheric components (see Figure \ref{impurities}), especially nitrogen
and oxygen, can be treated as inert gases that do not strongly affect
attachment kinetics. If nitrogen, oxygen, and the noble gases are all
effectively inert, then we are left with the hypothesis that trace active
impurities are likely responsible for the large dependence of $\alpha $ on
background gas pressure. Assuming that an equilibrium is established between
gaseous and surface impurities, we saw above that the surface impurity
density is expected to scale as $\rho _{surf}\sim P.$ Since $P$ spans a
factor of 200 in our measurements, we believe that impurities could account
for the observations shown in Figure \ref{lowp}.

\subsubsection{Hysteresis Effects}

Our suspicions regarding surface impurities were strengthened with the
further discovery of hysteresis effects in ice growth at 0.005 bar \cite%
{hysteresis}, as shown schematically in Figure \ref{theory}. In these
measurements we found that slowly increasing the growth of an ice crystal
(by slowly increasing the supersaturation) often resulted in a sudden growth
"surge" that would suddenly increase $\alpha $ by a factor of four or more.
After such a growth increase, $\alpha (\sigma )$ remained large as long as
the crystal continued growing, as shown in Figure \ref{theory}. If the
growth was halted for approximately 5-10 minutes $(\sigma \rightarrow 0)$,
then $\alpha \left( \sigma \right) $ was reduced to lower values.

The most reasonable explanation of this hysteresis phenomenon is that a
period of rapid growth drives off many of the impurities on an ice surface.
Assuming that impurities decrease $\alpha ,$ removing impurities can result
in an instability that explains the growth surge \cite{hysteresis}. The
growth remains rapid as long as impurities are not allowed to build up again
on the surface. Halting the growth allows the surface impurity density to
increase, which then reduces $\alpha (\sigma )$ to that of a dirty surface.
Rapid growth in near vacuum is then seen as a method for cleaning the ice
surface, after which one can measure $\alpha (\sigma )$ for an
uncontaminated surface. The top curve in Figure \ref{lowp} represents the
approximate locus of the maximum \textquotedblleft
post-surge\textquotedblright\ growth rates we measured \cite{hysteresis},
which we interpret as being the normal growth of a clean ice surface.

With all these observations, our hypothesis of surface impurities reducing
the ice growth rate provides a natural interpretation for the most striking
features in our data, whereas we can find no other ready explanation.

\begin{figure}[t] 
  \centering
  \includegraphics[width=6in,keepaspectratio]{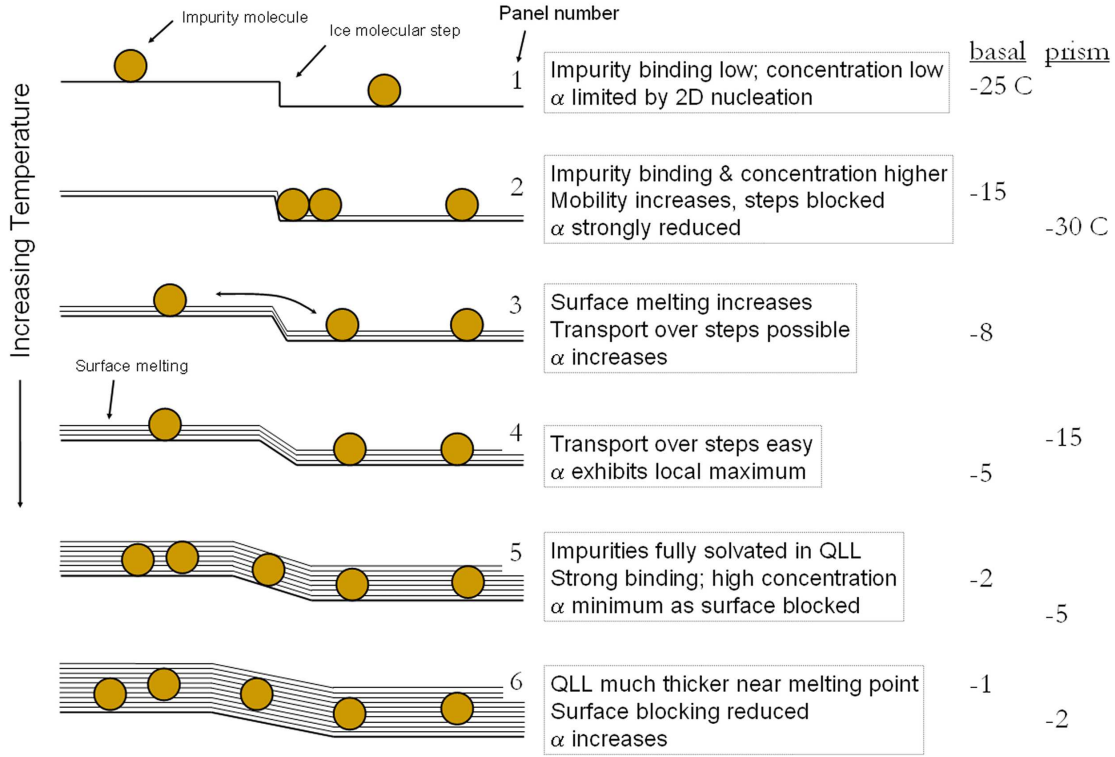}
  \caption{Our model for ice crystal growth
in the presence of surface melting and surface impurities. Different panels
refer to different temperatures, with colder temperatures at the top. The
columns on the right show temperatures for each growth behavior on the basal
and prism facets. See text for details.}
  \label{model}
\end{figure}

\subsubsection{Impurities and Surface Melting}

To better understand the ice growth data, it is necessary to develop a
molecular model that describes how the impurity effects vary with
temperature and with facet surface. As a first step in this direction, we
have constructed the model shown in Figure \ref{model}. Referring to the
panels in this figure, starting at the top, our model has the following
features:

\textbf{Panel 1)} At low temperatures, surface melting is negligible, so the
ice surface is essentially rigid and static. There is little
\textquotedblleft smoothing\textquotedblright\ of molecular steps by surface
melting, so the edge free energy $\beta $ is relatively high \cite{principal}%
. Impurity molecules have a relatively weak binding to the rigid ice
surface, so the equilibrium impurity surface density $\rho _{surf}$ is low.
In this regime, impurities have relatively little effect on the crystal
growth rates, which are then mainly limited by 2D nucleation as on a clean
ice surface.

\textbf{Panel 2)} As the temperature increases, surface melting begins to
produce some disorder at the interface, represented by thin lines in the
Figure. This disorder begins to smooth the molecular steps and reduce $\beta
,$ which is represented by a sloped molecular step in the Figure. The
increased mobility of water molecules also results in a partial solvation of
impurity molecules, which increases their binding to the surface, thus
increasing $\rho _{surf}$. The impurity molecules exhibit increased mobility
on the partially melted ice surface, and they bind more readily to steps
than to the molecularly flat faceted regions. When an ice island forms via
2D nucleation and grows in extent, the moving step pushes impurity molecules
along so they build up at the step edge, where they block subsequent growth.
As a result, $\alpha $ is strongly reduced.

\textbf{Panel 3)} At still higher temperatures, surface melting increases
and $\beta $ is reduced. The increased agitation and disorder of water
molecules on the surface increases the mobility of impurity molecules, which
can now sometimes hop over the smoothed steps. This tends to reduce the
buildup of impurities at the steps, so they do not block so completely the
growth of islands. Thus $\alpha $ increases in comparison to Panel 2.

\textbf{Panel 4)} Again, increased surface melting leads to greater impurity
mobility and an effectively reduced step height, so impurities now readily
move over steps. Island growth is not blocked and $\alpha $ exhibits a local
maximum.

\textbf{Panel 5)} The QLL now becomes thick enough that impurity molecules
are essentially completely solvated, which greatly increases their binding
to the surface. The impurity density increases to the point that the surface
is now coated with a dense impurity layer. This blocks access to the
underlying ice, which reduces $\alpha .$

\textbf{Panel 6)} Surface melting diverges rapidly near the melting point,
so now impurity molecules are essentially \textquotedblleft
floating\textquotedblright\ in the QLL. The layer is so thick that it takes
a time much greater than $\tau _{mono}$ for the impurity concentration to
build up to where it will substantially block ice growth. As the blocking
timescale becomes comparable to or longer than the growth time, $\alpha $
increases.

On the right side of Figure \ref{model}, we have indicated temperatures at
which the different growth behaviors are exhibited by the basal and prism
facet surfaces in our model. There is a systematic offset in the temperature
behavior of the two facets, which stems from an assumption that surface
melting is generally more pronounced on the prism facet in comparison to the
basal facet at a given temperature. A similar assumption was made in the LSK
model, but it is neither excluded nor supported by our existing knowledge of
surface melting in ice. It is also possible that there are differences in
impurity binding and impurity mobility on the two facet surfaces, and these
may also contribute to differences in the growth behavior with temperature.

This model is both complex and quite speculative, with only modest
theoretical and experimental backing. We expect that some details will
require modification and further refinement, but we believe that the
overarching concepts are sound. In particular, we believe that surface
impurities strongly reduce ice growth in comparison to that of an
impurity-free surface, and that the complex interplay of surface melting and
surface impurities is largely responsible for determining ice growth rates.
Before examining some of the testable predictions this model makes, we first
examine some of its features and describe how it can explain the snow
crystal morphology diagram.

\subsubsection{2D Nucleation}

One of the principal effects of impurities in our model is to reduce the
surface transport of water molecules to the edges of growing 2D islands, for
example in Panel 2 of Figure \ref{model}. We believe that surface impurities
do not act as nucleation centers to initiate the growth of islands, nor do
they substantially change the step free energy. Thus impurities do not
greatly affect the 2D nucleation process in general. If we parameterize the
attachment coefficient as $\alpha =A\exp (-\sigma _{0}/\sigma ),$ which is
characteristic for 2D nucleation-limited growth \cite{libbrechtreview}, then
our model suggests that the addition of surface impurities would change $A$
a great deal while having little effect on $\sigma _{0}.$ Indeed, we have
observed in all our low-pressure growth data \cite{principal, precision,
hysteresis} that $\sigma _{0}$ is remarkably insensitive to experimental
parameters at a fixed temperature, in agreement with our model.

At the higher temperatures shown in Figure \ref{model}, we believe that $%
\sigma _{0}$ is small enough that 2D nucleation no longer limits growth to a
great extent. Instead, we believe that the dominant growth-limiting factor
is from impurities directly blocking the growing surface, as described above.

\subsubsection{Kinetic Surface Melting}

At high growth rates we expect that the effective thickness of the QLL would
increase relative to its equilibrium value, which we refer to as \textit{%
kinetic surface melting}, corresponding to the more familiar kinetic
roughening \cite{saito}. Changes in the QLL thickness will in turn affect
the surface dynamics in Figure \ref{model}, which will alter growth rates.
The various effects of kinetic surface melting may be observable as changes
in growth behavior with supersaturation.

\begin{figure}[t] 
  \centering
  \includegraphics[width=5in,height=3.73in,keepaspectratio]{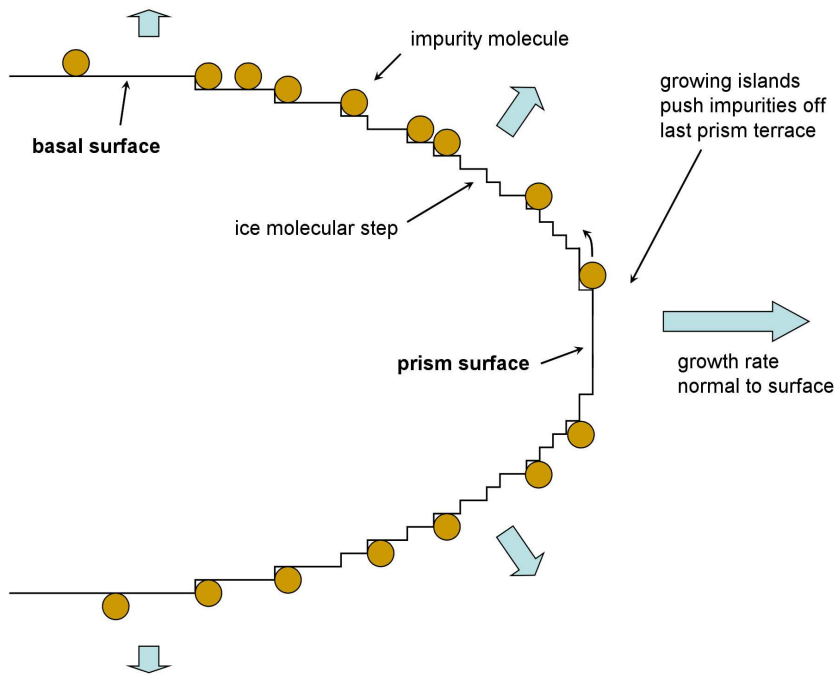}
  \caption{An edge-sharpening growth
instability (ESGI), as demonstrated for the growth of a thin plate-like
crystal (shown here in cross section near a growing edge). Because the plate
is thin, the width of the last ice terrace can be of order 100 molecular
diameters (see text). The growth of islands on this narrow terrace tends to
push impurity molecules to the side, thus cleaning the terrace. With fewer
impurities, $\protect\alpha $ increases on the prism surface, and so does
the growth rate. This effect results in a growth instability that sharpens
the edges of growing plate-like crystals.}
  \label{edge1}
\end{figure}

\subsubsection{An Edge-sharpening Growth Instability}

We believe that the intrinsic changes in $\alpha $ described above are
effectively amplified by an additional edge-sharpening growth instability
(ESGI), which is necessary to explain the growth of thin plates and other
features in the morphology diagram \cite{edge}. In the model presented here,
this instability comes about through impurity migration away from narrow
terraces, as shown schematically in Figure \ref{edge1}. For growth at -15 C,
the radius of curvature at the edge of a thin plate is of order $R\approx 1$ 
$\mu $m. The width of the last molecular terrace on the prism surface is
then $w\approx \sqrt{8aR}\approx 100a,$ where $a$ is the size of a water
molecule. As new islands form via 2D nucleation on top of this terrace, they
will tend to push impurity molecules laterally, removing impurities from the
terrace, which increases $\alpha _{prism}$ as shown in Figure \ref{edge1}.

This behavior produces an edge-sharpening growth instability because the
resulting $\alpha _{prism}$ depends on $R.$ If we begin with a broad terrace
(at the edge of a thick plate crystal), then impurity migration will clean
the terrace somewhat, which will enhance the edge growth and result in a
slightly smaller $R.$ This reduces $w,$ leading to more efficient cleaning
of the terrace, and thus still higher $\alpha .$ In addition, the faster
prism growth will give impurities less time to build up from the surrounding
gas. Both effects result in a positive feedback that leads to thinner plates
with faster edge growth. Kinetic surface melting may also play a role in
this instability when the supersaturation is high.

In our diffusion modeling of thin plate growth, we extract $\alpha _{prism}$
from the growth velocity of the prism surface, which is determined by the
growth rate on the last terrace. We believe that the high $\alpha _{prism}$
shown in Figure \ref{alphacomb} at -15 C results mainly because the prism
surface has been cleaned by the ESGI. For a large prism facet at this same
temperature, there would be no cleaning of the surface, so $\alpha _{prism}$
one would much lower. The ESGI can greatly amplify small intrinsic
differences in $\alpha $ between the two facets to produce the much larger
differences seen in Figure \ref{alphacomb}. As we pointed out earlier \cite%
{edge}, this factor must be taken into consideration when comparing growth
models for large, flat surfaces (Figure \ref{model}) with numbers extracted
from the growth of plate-like or needle-like crystals (Figure \ref{alphacomb}%
). Amplification via the ESGI is most substantial for the growth of thin
plates (near -15 C and - 2 C) and slender needles and sheaths (near -5 C and
below -30 C \cite{fieldguide, cold}).

\subsubsection{The Morphology Diagram}

Having laid out the characteristics of our molecular growth model, we can
now examine ice growth at different temperatures to explain the macroscopic
characteristics of the snow crystal morphology diagram. We consider
different temperature regions separately.

\textbf{Near -15 C.} At this temperature the growth of the basal facet in
air is strongly suppressed (Panel 2 in Figure \ref{model}), giving $\alpha
_{basal}\approx 0.001$ when $\sigma _{surface}$ is a few percent. This must
be a robust feature in any ice growth model, as thin, plate-like growth is a
well-established feature of snow crystal growth in air (and other inert
gases) at one bar. For growth at low pressures, however, the impurity
surface density is much lower, so $\alpha _{basal}$ is much higher,
explaining the data in Figure \ref{lowp}.

The intrinsic $\alpha _{prism}$ for a large, flat surface is higher than $%
\alpha _{basal}$ in air at -15 C, but it is not a great deal higher (Panel
4). Because $\alpha _{basal}$ is so low, however, the edge-sharpening growth
instability described above quickly takes effect on the prism facet. The
ESGI cleans the prism surface and greatly increases $\alpha _{prism}$
relative to what it would be on a large, flat surface, again explaining the $%
\alpha _{prism}$ data in Figure \ref{alphacomb} at -15 C.

At temperatures somewhat above or below -15 C, $\alpha _{basal}$ is higher
than the value at -15 C (panels 1 and 3), yielding a local minimum in $%
\alpha _{basal}$ around -15 C, as seen in Figure \ref{alphacomb}. The ESGI
is then less effective at cleaning the prism facet, which yields a local
maximum in $\alpha _{prism}$ near -15 C. Note that the growth of very thin
plates in air, and the large effective $\alpha _{prism}$ values needed,
results mainly from the removal of impurities from the prism facet by the
ESGI.

\textbf{Near -5 C.} At this temperature the growth of the prism surface is
suppressed by strong binding in the QLL (Panel 5), while the basal facet
growth is substantially less suppressed (Panel 4), resulting in columnar
growth. Under growth conditions with fairly low $\sigma ,$ the ESGI is not
present, so columns grow with blunt ends. At sufficiently high $\sigma ,$
however, the ESGI can take effect on the basal facets, resulting in the
growth of sheath-like crystals \cite{fieldguide, cold}.

\textbf{Near -2 C.} Here basal growth is suppressed by strong binding in the
QLL\ (Panel 5), while the QLL\ is so thick on the prism facet that its
growth is increased relative to that at -5 C. Because $\alpha _{basal}$ is
low, the ESGI can take effect to increase $\alpha _{prism}$ to the value
shown in Figure \ref{alphacomb} and produce thin, plate-like crystals. At
still higher temperatures, the growth of both facet surfaces is at some
intermediate value (Panel 6), so the ESGI is no longer present, resulting in
a more isometric morphology.

\textbf{Below -30 C.} Our model is most uncertain at low temperatures, but
we expect growth of the prism surface will be suppressed (Panel 2), leading
to the growth of columnar crystals. The ESGI will then lead to the growth of
sheath-like crystals at high supersaturations, as observed \cite{fieldguide,
cold}.

\section{Discussion}

We are enthusiastic about this new ice growth model for a number of reasons.
First, it fairly naturally explains many previously puzzling features in our
ice growth data, especially the substantial variation in the attachment
coefficient with background gas pressure. Second, the model provides a
viable explanation of the snow crystal morphology diagram, which has baffled
researchers for over 60 years. Third, the influence of low levels of
impurities is likely responsible for some of the disparity seen in previous
ice growth experiments \cite{critical}. Fourth, the model is easily testable
with additional growth experiments. And finally, the notion of impurities
playing an essential role in normal ice growth has been largely ignored to
date, so our model opens up new avenues for exploring impurity-dependent
growth behavior.

Verifying all the microscope details outlined in Figure \ref{model} will be
an extremely difficult task, involving precise measurements of the
characteristics of ice surface melting, impurity attachment kinetics, as
well as impurity binding and mobility, all as a function of temperature. In
addition, we would need to understand how these surface characteristics
affect the molecular dynamics of crystal growth if we wish to determine
growth rates and morphologies. Direct molecular-scale imaging of the ice
surface, such as with scanning probe microscopy, is of limited use because
the molecular motions are extremely rapid and the equilibrium vapor pressure
is high, making the ice surface too unstable for most molecular-scale
imaging techniques. Molecular dynamics simulations are also challenging for
such complex surfaces.

Fortunately, one can use crystal growth itself as a diagnostic tool for
exploring the molecular dynamics of the ice surface. By making additional,
relatively straightforward measurements of crystal growth rates and
morphologies as a function of temperature, supersaturation, background gas
pressure, gas constituents, and surface orientation relative to the lattice,
one should be able examine a great many characteristics of our model in
considerable detail. Although our model is quite speculative at present, it
provides a useful guide for additional experiments that will no doubt lead
to modifications and additional refinements.

Some potentially interesting directions include:

1) In extremely clean air (or other inert gases) at one bar, the growth
rates should be very high because the ice surface would be free from
impurities. The ESGI would not be present, and the morphology diagram would
look very different from Figure \ref{morph}. This would provide a dramatic
\textquotedblleft smoking gun\textquotedblright\ verification of the most
important feature in our model, namely that even low levels of impurities
substantially reduce ice growth rates. Unfortunately, creating an
impurity-free gas is an experimental impossibility, and we cannot accurately
estimate at present what impurity levels are necessary to produce growth
rates that approach that of a clean ice surface. It appears likely, however,
that even modest cleaning efforts will yield some measurable changes.

2) Even if one can only increase impurity levels relative to those found in
ordinary laboratory air, measuring growth rates as a function of impurity
density should provide many valuable insights into our model. High impurity
levels would change many of the details in Figure \ref{model} and would
significantly alter the behavior of the ESGI. Experiments of this nature
have not yet been done, but many should be quite straightforward.

3) We have little knowledge at present about which impurities have the
greatest effect on ice growth, and which are most important under normal
atmospheric conditions. Exploring the chemistry of impurity-mediated ice
crystal growth should shed considerable light on the microscopic mechanisms
involved.

Although snow crystal growth is a seemingly simple process, a closer look
reveals that the underlying physics and chemistry are remarkably subtle and
interesting. The observed morphological diversity of snow crystals results
from a complex interplay of molecular dynamical processes occurring at the
ice surface. There is much yet to be learned by investigating this
fascinating phenomenon.

\end{document}